# Models of Synthesis of Uniform Colloids and Nanocrystals

**Vyacheslav Gorshkov** [a, b]  and  **Vladimir Privman** [c]

[a] Institute of Physics, National Academy of Sciences, 46 Nauky Avenue, Kiev 680028, Ukraine

[b] National Technical University of Ukraine "KPI," 37 Peremogy Avenue, Building 7, Kiev-56, 03056 Ukraine

[c] Center for Advanced Materials Processing, Department of Physics, Clarkson University, Potsdam, NY 13699, USA

**Abstract**

We present modeling approaches to explain mechanisms of control of uniformity (narrow distribution) of sizes and shapes in synthesis of nanosize crystals and micron-size colloids. We consider those situations when the nanocrystals are formed by burst nucleation. The colloids are then self-assembled by aggregation of nanocrystals. The coupled kinetic processes are both controlled by diffusional transport, yielding well-defined colloid dispersions used in many applications. We address aspects of modeling of particle structure selection, ranging from nucleation to growth by aggregation and to mechanisms of emergence of particle shapes.

# 1. Introduction

Numerous applications and scientific studies require the use of "uniform" colloids and nanoparticles. Mechanisms for obtaining particles of narrow size distribution and limited variation of shapes, can be different depending on the situation and type of the particles. Colloids are suspensions of few-micron down to sub-micron size particles, whereas nanoparticles are of much smaller sizes, of order 0.01 µm (10 nm) and smaller. Synthesis of well-defined products has to address an even broader goal of achieving uniformity in particle composition, internal structure and morphology, and surface structure.

A theoretical modeling program has to include understanding of and numerical-simulation approaches to processes of nucleation, growth, aggregation, and surface interactions of fine particles. Here we review our modeling work and numerical approaches for studying [1-15] burst-nucleation of crystalline nanoparticles in solution, the accompanying process of diffusional aggregation of these nanoparticles to form uniform polycrystalline colloids, and considerations of shape selection in particle growth.

For applications, colloids are usually considered "monodispersed" for narrow particle diameter distributions, of relative width 6-12%. For nanosize particles, it is expected that for certain nanotechnology device applications the requirements are event stricter: uniform size and shape may imply "atomically identical." Therefore, experimental methodologies for controlling size and shape distributions, which have a long history, e.g., [16,17], in connection with applications of colloid suspensions, have become more important and timely with the advent of nanotechnology.

We consider the "building blocks" from which particle are formed, as well as particles themselves, suspended and transported by diffusion in solution. The former, "monomers" or "singlets," in nanoparticle synthesis are atomic-size solute species: atoms, ions, molecules, whereas for colloids, they are the nanosize, typically nanocrystalline primary particles. In the colloid case, the supply of singlets is controlled by their own burst nucleation. However, the monomers for both processes can also be added externally.

A desirable particle size distribution, with a relatively narrow peak at large cluster sizes, is shown in Figure 1. We note that cluster-cluster aggregation or cluster ripening due to exchange of monomers—examples processes that make the size distribution grow—also broaden it. They cannot lead to narrow peak formation. Most growth/coarsening mechanisms broaden the distribution because larger particles have bigger surface area for capturing "building blocks," as well as, e.g., for spherical shapes, less surface curvature, which implies generally slightly better binding and less detachment of monomers.

One approach to getting a narrow size/shape distribution has been by actually blocking the growth of the "right side" of the main peak in Figure 1, by caging the growing clusters inside nanoporous structures or objects (such as micelles or inverse micelles). This technique was reviewed, e.g., in [18]. It has a disadvantage of requiring the use of additional chemicals that later remain part of the formed particles.

Another templating approach to achieve desired material compositions and shapes/morphologies is by growth by deposition on top of seeded earlier-prepared smaller uniform size and shape particles. It is reviewed, e.g., in [19].

Burst nucleation, defined and studied [9-11] in Section 2, demonstrates another size-selection approach whereby the left side of the peak is eroded fast enough as compared to the peak broadening by coarsening processes, to maintain narrow distribution. In practice, other coarsening processes eventually broaden the distribution after the initial nucleation burst. Thus, this approach at best works for nano-sizes.

In Section 3, we introduce an important mechanism [1,6,10,11] for attaining *colloid* particle size distributions narrow on a *relative* scale. This involves a large supply of monomers, of concentration $C(t)$, see Figure 1. The monomers "feed" the peak, thus pushing it to larger sizes, and the process can be fast enough not to significantly broaden the central peak, and, with a proper control of $C(t)$, not to generate a significant "shoulder" at small clusters. The singlets for such a process yielding colloids, are actually the burst-nucleated nanocrystalline precursors (primary particle). For nanosize particle preparation, there has also been interest in stepwise processes, e.g., [20,21], with batches of atomic-size "monomers" supplied to induce further growth of the earlier formed nanoparticles.

To be somewhat more specific, let $N_s(t)$ denote the density of particles containing $s$ singlets, at time $t$. The size distribution evolves in time with a peak eventually present at some relatively large $s$ values, see Figure 1. Let us denote the singlet concentration by

$$C(t) \equiv N_1(t).  \tag{1}$$

The singlets can be supplied as a batch, several batches, or at the rate $\rho(t)$ (per unit volume). They are consumed by the processes involving the production of small clusters, in the "shoulder" in Figure 1. They are also consumed by the growing large clusters in the main peak.

There are several issues to consider in uniform particle growth. First, how is the main peak formed in the first place? Second, and most important, how to grow the main peak without much broadening? What are the roles of various processes such as cluster-cluster aggregation, etc.? In Section 4, we continue the discussion of Section 3 to address some of these issues and describe elaborations of the initially introduced model.

In nanoparticle synthesis the main mechanism of the early formation of the peak is by burst nucleation, when nuclei of sizes larger than the critical size form from smaller-cluster "embryos" by growing over the nucleation barrier. Of course, *seeding* is another way of initiating the peaked size distribution both for colloid and nanoparticle growth. For colloid synthesis without seeding, the initial peak formation can be driven by the supply of singlets and/or can also be facilitated by cluster-cluster aggregation at the early growth stages.

Finally, the preceding discussion of the size distribution did not address the problem of *shapes* (and more generally, morphology) in nanoparticle and colloid synthesis. As discussed in Section 5, this is presently a largely open problem: Several mechanisms for particle shape selection in fast, nonequilibrium growth have been considered, and probably some or all are valid depending on the details of the system. For uniform-shape growth, we have advanced a thesis that fast growth without development of large internal defect structures can lead to shape selection with non-spherical particle "faces" similar to those obtained in equilibrium crystal structures (but of different proportions). This approach is presented in Section 5. Finally, Section 6 offers concluding comments.

## 2. Burst nucleation

The model of burst nucleation [9-11,22,23] is appropriate for growth of nanosize particles, consisting of $n$ monomers. Particles with $n > n_c$, where $n_c$ is the critical cluster size (to be defined shortly), irreversibly capture diffusing solutes: atoms, ions or molecules. The dynamics in the shoulder, for $n < n_c$, see Figure 2, is such that the subcritical ($n < n_c$) "embryos" are assumed instantaneously rethermalized.

Burst nucleation occurs in a supersaturated solution with time-dependent monomer concentration $c(t)$. As typical for nucleation theory approaches, we assume that thermal fluctuations cause formation of small embryos. This process is controlled by the free-energy barrier imposed by the surface free energy. The kinetics of these few-atom-size clusters involves complicated transitions between embryos of various sizes, shapes, as well as internal restructuring. These processes are presently not well understood. However, the dynamics is fast, and one can assume that cluster sizes are approximately thermally distributed and controlled by a Gibbs-like form of the free energy of an *n*-monomer embryo,

$$\Delta G(n,c) = -(n-1)kT \ln(c/c_0) + 4\pi a^2 \left(n^{2/3} - 1\right)\sigma. \tag{2}$$

Here $k$ is the Boltzmann constant, $T$ is the temperature, $c_0$ is the equilibrium concentration of monomers, and $\sigma$ is the effective surface tension. This expression increases with $n$ until it reaches the maximum value at the nucleation barrier, attained at $n_c$,

$$n_c(c) = \left[\frac{8\pi a^2 \sigma}{3kT \ln(c/c_0)}\right]^3. \tag{3}$$

For $n > n_c$, the free energy decreases with $n$, but the kinetics of such clusters is irreversible and is not controlled by free-energy considerations.

The first term in Equation (2) is the bulk contribution. It is derived from the entropy of mixing of noninteracting solutes and is negative for $c > c_0$, therefore favoring formation of large clusters. The second, positive term represents the surface free-energy, proportional to the area, $\sim n^{2/3}$. This term dominates for $n < n_c$, and its competition with the bulk term results in the presence of the nucleation barrier. The effective solute radius, $a$, is defined in such a way that the radius of an $n$-solute embryo is $an^{1/3}$. It can be estimated by requiring that $4\pi a^3/3$ equals the unit-cell volume per singlet (including the surrounding void volume) in the bulk material.

The approach here is the same as in most treatments of homogeneous nucleation, with the unique aspect being that the bulk free energy expression is dependent on the monomer concentration and therefore varies with time. As usual, we assume that the distribution of aggregate shapes can be neglected: a "representative" embryo is taken spherical in the calculation of its surface area and the monomer transport rate to it. We note that the surface tension of spherical particles varies with their size. This effect, as well as any other geometrical factors that might be needed because real clusters are not precisely spherical, is neglected. The effective surface tension of nanoparticles is only partially understood at present [24]. Thus, $\sigma$ was either assumed [1,5,7,8] close to $\sigma_{\text{bulk}}$ (which might not always be correct for particles smaller than 5-10 nm), or fitted as an adjustable parameter.

Significant suppression of nucleation occurs after the initial burst, during which $c/c_0$ decreases from the initial value $c(0)/c_0 \gg 1$ towards its asymptotic large-time equilibrium value 1. The large-time form [9-11] of the particle size distribution in burst nucleation is shown in Figure 2. Specifically, embryos smaller than $n_c$ are thermalized on time scales much faster than those of other dynamical processes. Their size distribution is

$$P(n<n_c,t) = c(t)\exp\left[\frac{-\Delta G(n,c(t))}{kT}\right], \tag{4}$$

defined so that the concentration (per unit volume) of embryos with sizes in $dn$ is $P(n,t)dn$. Here $n_c = n_c(c(t))$.

The rate of production of supercritical clusters, $\rho(t)$, can be written [1] as follows,

$$\rho(t) = K_{n_c} cP(n_c,t) = K_{n_c} c^2 \exp\left[\frac{-\Delta G(n_c,c)}{kT}\right], \tag{5}$$

where

$$K_n = 4\pi a n^{1/3} D, \tag{6}$$

is the Smoluchowski rate [2,25] for the irreversible capture of diffusing solutes by growing spherical clusters of sizes assumed $n \geq n_c \gg 1$. here $D$ is the diffusion coefficient for monomers in a dilute solution of viscosity $\eta$; $D$ can be estimated as $\sim kT/6\pi\eta a$, up to geometrical factors (the effective radius $a$ should be related to the hydrodynamic radius).

Rapid growth of the supercritical, $n > n_c$, clusters can be modeled [9] by using the kinetic equation

$$\frac{\partial P(n,t)}{\partial t} = (c(t) - c_0)(K_{n-1} P(n-1,t) - K_n P(n,t)), \tag{7}$$

where the difference $c(t) - c_0$ is used here in place of $c(t)$ to ensure that the growth of clusters stops as $c(t)$ approaches $c_0$. This factor approximately accounts [2] for detachment of matter if we ignore curvature and similar effects. Specifically, variation of the nanocluster surface tension with its radius is accompanied by a variation of the effective "equilibrium concentration" and gives rise to Ostwald ripening [26] driven by exchange of monomers. This and other possible coarsening processes, such as cluster-cluster aggregation [27,28], are neglected here because burst nucleation is expected [1,9-11] to be initially a fast process. However, for later times these additional coarsening processes will gradually widen (while further growing) the particle size distribution, which for burst nucleation alone is well characterized by the function $n_c(t)$ sketched in Figure 2, to be quantified shortly. Indeed, it turns out that for most systems the large-time linear growth of $n_c(t)$, see Figure 2, has a very small slope [29]: the growth would practically stop if it were due to burst-nucleation alone.

In addition to growth/shrinkage by attachment/detachment of matter, particles of all sizes also undergo internal restructuring, modeling of which for nanosize clusters is still not well developed [30,31]. Without such restructuring, the clusters would grow as fractals [27,28], whereas density measurements and X-ray diffraction data for colloidal particles aggregated from burst-nucleated nanosize subunits indicate that they have polycrystalline structure and density close to that the bulk [1,16]. There is primarily experimental, but also modeling evidence [1,4,5,7,8], that for larger clusters internal restructuring leads to compact particles with smooth surfaces, which then grow largely irreversibly.

The supercritical distribution, $n > n_c(t)$, see Figure 2, irreversibly grows by capturing monomers, but, at the same time, the subcritical, $n < n_c(t)$, matter is redistributed by fast thermalization. The function $n_c(t)$ is increasing monotonically. Obviously, his sharp cutoff between two types of dynamics is an approximation, typical of nucleation theories. The short-time form of the supercritical distribution depends on the initial conditions. At large times [9-11] it will eventually have its maximum at $n = n_c$, and will take on the form of a truncated Gaussian: the peak of the full Gaussian curve, only the "right-size slope" of which is shown in Figure 2, is actually to the left of $n_c$.

These expectations were confirmed by extensive numerical modeling of time-dependent distributions, for several initial conditions, obtained by a novel efficient numerical integration scheme which is not reviewed here; see [9]. In what follows, we concentrate on the derivation of analytical results for large times. It can be shown that the kinetic equation has an asymptotic solution of the Gaussian form

$$P_G(n,t) = \zeta(t) c_0 \exp\left[-(\alpha(t))^2 (n - K(t))^2\right], \tag{8}$$

for $n > n_c(t)$ and large $t$. The "peak offset" $n_c(t) - K(t)$ is a positive quantity. The derivation starts with writing Equation (7) in a continuous-$n$ form, keeping terms up to the second derivative, in order to retain the diffusive nature of the peak broadening,

$$\frac{\partial P}{\partial t} = (c - c_0)\left[\left(\frac{1}{2}\frac{\partial^2}{\partial n^2} - \frac{\partial}{\partial n}\right)(K_n P)\right]. \tag{9}$$

Since irreversible growth of supercritical clusters corresponds to $P(n,t)$ taking on appreciable values only over a narrow range, we can further approximate $K_n \approx K_{n_c} = \kappa(n_c(t))^{1/3}/c_0$, where

$$\kappa \equiv 4\pi c_0 a D. \tag{10}$$

We define the dimensionless quantity

$$x(t) \equiv c(t)/c_0, \tag{11}$$

and utilize Equations (3) and (6) to show that the product of the coefficients, $(c - c_0)K_{n_c}$, becomes a constant in the limit of interest, to yield

$$\frac{\partial P}{\partial t} = \frac{z^2}{2}\left(\frac{1}{2}\frac{\partial^2}{\partial n^2} - \frac{\partial}{\partial n}\right)P, \tag{12}$$

where we conveniently defined

$$z^2 \equiv \frac{64\pi^2 a^3 \sigma c_0 D}{3kT}. \tag{13}$$

Equation (12) implies that the solution is indeed a Gaussian, with the parameters, introduced in Equation (8), given by

$$\alpha(t) \approx 1/\sqrt{z^2 t}, \quad K(t) \approx z^2 t/2, \quad \zeta(t) \approx \Omega/\sqrt{z^2 t}. \tag{14}$$

The prefactor $\Omega$ cannot be determined from the asymptotic analysis alone, because the overall height of the distribution is obviously expected to depend on the initial conditions. It, and certain other quantities, have to be determined from the conservation of matter. Specifically, for the "peak offset," rather complicated mathematical considerations, presented in [9], yields the result that $n_c(t) - K(t) \propto \sqrt{t \ln t}$ (with a positive coefficient). Since $K(t)$ growth linearly with time, this difference is then sub-leading and we finally the key result that

$$n_c(t) \approx z^2 t/2. \tag{15}$$

We also note that the width of the truncated Gaussian is given by $1/\alpha \propto \sqrt{t}$. Thus the *relative* width decreases with time according to $\sim t^{-1/2}$. One can also confirm [9] that the difference $c(t) - c_0$ approaches zero ($\sim t^{-1/3}$).

The Gaussian distribution has also provided a good fit at intermediate times for numerical data for various initial conditions, including for initially seeded distributions; see [9]. Numerical simulations also reveal the other expected features of burst nucleation, summarized in Figure 2: The initial induction period followed by growth "burst" that precedes the onset of the asymptotically linear growth. Experimentally, it has been challenging to quantify distribution of nucleated nanocrystals, because of their tendency to aggregate and non-spherical shapes. The distribution is typically two-sided around the peak, and the final particles stop growing after a certain time. Both of these properties are at odds with the predictions of the burst-nucleation model, and the discrepancy can be attributed to the assumed instantaneous thermalization of the clusters below the critical size and to the role of other growth mechanisms.

Specifically, for very small clusters, below a certain cutoff value, which has been tentatively estimated [6-8,32-34] to correspond to $n_{th} \approx 15\text{-}25$ "monomers" (atoms, ions, molecules, sub-clusters), clusters can evolve very rapidly, so that the assumption of fast, thermalization/restructuring is justified. For larger sizes, embryos will develop a bulk-like core and their dynamics will slow down: once $n_c(t) > n_{th}$, the "classical" nucleation model should be regarded as approximate. Modifications of the model have been contemplated [9,35,36]. These, however, require system parameters which are not as well defined and as natural as those of the "classical" model. One of the interesting applications of the present model would be to estimate the deviations from the "classical" behavior and thus the value of $n_{\text{th}}$ — the nanocluster size beyond which a "bulk-material" core develops.

## 3. Colloid synthesis

Burst nucleation, which ideally can yield narrow size distributions, can yield particles up to several tens of nanometers in diameter. Size distribution of particles nucleated in the initial burst is then usually broadened as they further grow by other mechanisms. However, one notable exception exists: the two-stage mechanism [1] whereby the nanosized primary particles, burst-nucleated and growing in solution, themselves become the singlets for the aggregation process which results in uniform secondary particles of colloid dimensions, from submicron to few microns in diameter.

Many dispersions of uniform colloid particles of various chemical compositions and shapes, have been synthesized with their structural properties consistent with such a two-stage mechanism [1,7-8,16,37-60]. Specifically, spherical particles precipitated from solution showed polycrystalline X-ray characteristics, such as ZnS [39], CdS [7,8,38], $Fe_2O_3$ [37], Au and other metals [1,23,54-56,58,60]. Furthermore, experimental techniques have confirmed that these and many other monodispersed inorganic colloids consist of nanocrystalline subunits [1,7-8,16,37-56,58-60]. It was observed [1,52,54] that these subunits were of the same size as the diameter of the precursor singlets of sizes of order up to a couple of 10 nm, formed in solution, thus suggesting an aggregation mechanism. This two-stage growth process is summarized in Figure 3. The composite structure has also been identified for some uniform *nonspherical* colloid particles [37,46,48,50,59], but the findings are not definitive enough to commit to the specific two-stage growth mechanism discussed here.

We first consider a model that involves the coupled primary and secondary processes in the simplest possible formulation that involves various approximations but avoids introduction of unknown microscopic parameters. In the next section, we describe certain improvements that allow for a better agreement with experimental observations. The latter approach, however, involves rather demanding numerical simulations, and therefore details of the computational aspects of the model equations are also presented. Additional information, examples of experimental parameters and results, and well as sample numerical data fits can be found in [1,5,7,8,12,13].

For the secondary particles, we assume growth by irreversible capture of singlets by the larger growing aggregates. This is particularly well suited to describe the evolution of the already well developed peak, see Figure 1, assuming that the role of the particles in the "shoulder" (Figure 1) is minimal. We then use rate equations, with $\Gamma_s$ denoting the rate constants for singlet capture by the $s \geq 1$ aggregates (to be quantified shortly), and all the other quantities defined earlier, in Section 1, in connection with Equation (1),

$$\frac{dN_s}{dt} = (\Gamma_{s-1}N_{s-1} - \Gamma_s N_s)C, \quad s > 2, \tag{16}$$

$$\frac{dN_2}{dt} = (\frac{1}{2}\Gamma_1 C - \Gamma_2 N_2)C, \tag{17}$$

$$\frac{dC}{dt} = \rho - \sum_{s=2}^{\infty} s \frac{dN_s}{dt} = \rho - \Gamma_1 C^2 - C \sum_{s=2}^{\infty} \Gamma_s N_s. \tag{18}$$

Thus we for now ignore cluster-cluster aggregation and for now assume that the only process involving the $s > 1$ aggregates is that of capturing singlets at the rate proportional to the concentration of the latter, $\Gamma_s C$, has been commonly used in the litrature, e.g., [1,5-6,61-63]. We will comment on elaborations later. More complex processes, such as cluster-cluster aggregation [27,28], detachment [2,4] and exchange of singlets (ripening), etc., also contribute to particle growth, and broaden the particle size distribution. However, in colloid synthesis they are much slower than the singlet-consumption driven growth.

An obvious approximation involved in writing Equation (16-18) is that of ignoring particle shape and morphology distribution. We dodge this issue, which is not well understood and difficult to model, by assuming that the growing aggregates rapidly restructure into compact bulk-like particles, of an approximately fixed shape, typically, but not always, spherical for colloids. This has been experimentally observed in uniform colloid synthesis [1,40-42,47,49,54]. Without such restructuring, the aggregates would be fractal [28,64]. We address shape selection in Section 5.

Generally for such "minimal" models of particle growth, if the singlets are supplied/available constantly, then the size distribution will develop a large shoulder at small aggregates, with no pronounced peak at $s \gg 1$. If the supply of singlets is limited, then only very small aggregates will be formed. A key discovery in studies of colloid synthesis [1,6] has been that there exist protocols of singlet supply, at the rate $\rho(t)$ which is a *slowly decaying function of time,* that yield peaked (at large sizes) distributions for large times. Furthermore, the primary — nanocrystal nucleation — process in uniform polycrystalline colloid synthesis, naturally "feeds" the secondary process — that of the nucleated nanoparticles aggregating to form colloids — just at a rate like this.

All the rates in the considered processes are diffusionally controlled. Specifically, diffusional growth of the secondary (colloid) particles is not always present and must be facilitated by the appropriate chemical conditions in the system: The ionic strength and pH must be maintained such that the surface potential is close to the isoelectric point, resulting in reduction of electrostatic barriers and promoting fast irreversible primary particle attachment [1,40-42,47,49,54]. The clusters of sizes *s*, present in solution with the volume densities $N_{s=1,2,3,...}(t)$, are be defined by how many primary particles (singlets) were aggregated into each secondary particle. For equations (16-18), we take the initial conditions $N_{s=1,2,3,...}(0) = 0$. The simplest choice of the rate constants is the Smoluchowski expression, encountered in Equation (6),

$$\Gamma_s \approx 4\pi R_p D_p s^{1/3}. \qquad (19)$$

Here $R_p$, $D_p$ are the effective primary particle radius and diffusion constant, the choice of which will be discussed shortly. The approximate sign is used because several possible improvement to this simplest formula can be offered, as will be described later. A typical numerical calculation result for a model of the type developed here is shown in Figure 4, illustrating the key feature — effective size selection — the "freezing" of the growth even for exponentially increasing times (here in steps $\times 10$).

For $\rho(t)$ in Equation (18), we will use the rate of production of the supercritical clusters, Equation (5), the calculation of which requires $c(t)$. We will use the following convenient [1] but approximate relation,

$$\frac{dc}{dt} = -n_c \rho, \qquad (20)$$

combined with Equations (3,5,6). Earlier we referenced [9], but not detailed complicated steps required in order to derive the expression (not shown here) for $dc/dt$ for burst nucleation alone, without the secondary aggregation process. When the burst-nucleated, growing supercritical particles are also consumed by the secondary aggregation, even more complicated considerations would be required. Indeed, the solute species (present with concentration $c$ in the dilute, supersaturated solution) are also partly stored in the $n > 1$ subcritical embryos, as well as in the supercritical primary particles and in the secondary aggregates. They can be captured by larger particles, as assumed in our model of burst nucleation, but they can also detach back into the solution.

The main virtue of the proposed approximation, Equation (20), is tractability. It basically ignores the effect of the possible rebalancing of the "recoverable" stored solute species in various part of the particle distributions, but rather it focuses on the loss of their availability due to the unrecoverable storage in secondary particles of sizes $s = 1, 2, 3, \ldots$ (where the $s = 1$ particles are the "singlet" nucleated supercritical nuclei, whereas $s > 2$ corresponds to their aggregates). The form of the right-hand side of Equation (20), when used with Equations (3,5,6), also ignores further capture by and detachment from larger particles. The resulting equations for calculating the rate $\rho(t)$ of the supply of singlets for the secondary aggregation, starting with the initial supercritical concentration $c(0) \gg c_0$ of solutes, are

$$\frac{dc}{dt} = -\frac{2^{14} \pi^5 a^9 \sigma^4 D_a c^2}{(3kT)^4 [\ln(c/c_0)]^4} \exp\left\{-\frac{2^8 \pi^3 a^6 \sigma^3}{(3kT)^3 [\ln(c/c_0)]^2}\right\}, \qquad (21)$$

$$\rho(t) = \frac{2^5 \pi^2 a^3 \sigma D_a c^2}{3kT \ln(c/c_0)} \exp\left\{-\frac{2^8 \pi^3 a^6 \sigma^3}{(3kT)^3 [\ln(c/c_0)]^2}\right\}, \qquad (22)$$

which can be used to numerically calculate $\rho(t)$. Here, we denoted the diffusion constant of the solutes by $D_a$, in order to distinguish it from $D_p$ of the primary particles. This completes the set of equations for the minimal model, yielding results of the type illustrated in Figure 4.

Let us now discuss the choice of parameters entering Equation (19), and some of the simplifying assumptions made in the formulated model. We will also consider, in the next section, possible modifications of the model. Figure 4, based on one of the sets of the parameter values used for modeling formation of uniform spherical Au particles, already includes some of these modification [5].

We note that if the assumption $s \gg 1$ is not made, the full Smoluchowski rate expression [2,25] should be used, which, for aggregation of particles of sizes $s_1$ and $s_2$, on encounters due to their diffusional motion, is

$$\Gamma_{s_1, s_2 \to s_1 + s_2} \simeq 4\pi \left[ R_p \left( s_1^{1/3} + s_2^{1/3} \right) \right] \left[ D_p \left( s_1^{-1/3} + s_2^{-1/3} \right) \right], \qquad (23)$$

where for singlet capture $s_1 = s$ and $s_2 = 1$. This relation can not only introduce nontrivial factors for small particle sizes, as compared to Equation (19), but it also contains an assumption that the diffusion constant of $s$-singlet, dense particles is inversely proportional to the radius, i.e., to $s^{-1/3}$, which might not be accurate for very small, few-singlet aggregates.

Another assumption in Equations (19,23) is that the radius of $s$-singlet, dense particles can be estimated as $R_p s^{1/3}$. However, primary particles actually have a distribution of radii, and they can also age (grow/coarsen) before their capture by and incorporation into the structure of the secondary particles. In order to partially compensate for this approximations, the following arguments can be used. Regarding the size distribution of the singlets, it has been argued that since their capture rate especially by the larger aggregates is proportional to their radius times their diffusion constant, this rate will not be that sensitive to the particle size and size distribution, because the diffusion constant for each particle is inversely proportional to its radius. Thus, the product is well approximated by a single typical value.

The assumption of ignoring the primary particle ageing, can be circumvented by using the experimentally determined typical primary particle linear size ("diameter"), $2R_{\exp}$, instead of attempting to estimate it as a function of time during the two-stage growth process. In fact, for the radius of the $s$-singlet particle, the expressions in the first factor in Equation (23), which represents the sum of such terms, $R_p s^{1/3}$, should be then recalculated with the replacement

$$R_p s^{1/3} \to 1.2 R_{\exp} s^{1/3}. \qquad (24)$$

The added factor is $(0.58)^{-1/3} \simeq 1.2$, where 0.58 is the filling factor of a random loose packing of spheres [65]. It was introduced to approximately account for that as the growing secondary particle compactifies by restructuring, not all its volume will be crystalline. A fraction will consists of amorphous "bridging regions" between the nanocrystalline subunits.

Finally, at the end of the model computations, inaccuracies due to the approximations entailed in using Equation (20), detailed earlier, and the use of the uniform singlet radii, Equation (24), both possibly leading to nonconservation of the total amount of matter, can be partly compensated for [1], by renormalizing the final distributions so that the particles per unit volume contain the correct amount of matter. This effect seems not to play a significant role in the dynamics. Some additional technical issues and details of the modeling are not reviewed here; see [1,3,5,7,8,12,13].

## 4. Improved models for colloid growth

Two-stage models of the type outlined in Section 3, were shown to provide a good semi-quantitative description (without adjustable parameters) of the processes of formation of spherical colloid-size particles of two metals: Au [1,3,5,7,12,13,66] and Ag [12,13,], a salt: CdS [7,8], as well as argued to qualitatively explain the synthesis of an organic colloid: monodispersed microspheres of Insulin [57].

Here we discuss additional elaborations, developed to improve the two-stage model of colloid synthesis to achieve quantitative agreement with experimental results, including size distributions of CdS [7,8], Au [66], and Ag [12,13] particles, the former measured for different times during the process and for several protocols of feeding the solutes into the system, rather than just for their instantaneous "batch" supply. For controlled release of ions, we have to include in the model the rate equations for their production in chemical reactions. This is in itself an interesting problem: identification and modeling of the kinetics of various possible intermediate solute species are not always well studied or understood theoretically, and they are not easy to probe experimentally.

In numerical simulations, the physical properties of the primary nucleation process: the values of the effective surface tension and of the equilibrium concentration, if not well-known experimentally and instead adjusted as fit parameters, were found to mostly affect the time scales of the secondary particle formation, i.e., the onset of "freezing" of their growth, illustrated in Figure 4. Accumulated evidence suggests that the use of the bulk surface tension and other experimentally determined parameters yields reasonable results consistent with the experimentally observed times.

The kinetic parameters of the secondary aggregation primarily control the size of the final particles. We found [1,3,5,7,8,12,13,66] that sizes numerically calculated within the "minimal" model, while of the correct order of magnitude, were smaller than the experimentally observed values, by a non-negligible factor, because the kinetics of the secondary aggregation results in too many secondary particles which, since the total supply of matter is fixed, then grow to sizes smaller than those experimentally observed.

Two modeling approaches to remedy this property were considered. The first argument [5,12] has been that for very small "secondary" aggregates, those consisting of one or few primary particles, the spherical-particle diffusional expressions for the rates, which are anyway ambiguous for tiny clusters, as described in connection with Equations (19,23,24), should be modified. Since the idea is to avoid introduction of many adjustable parameters, the rate $\Gamma_{1,1\to 2}$, cf. Equation (23), was multiplied by a "bottleneck" factor, $f<1$. Indeed, merging of two singlets (and other very small aggregates) may involve substantial restructuring, thus reducing the rate of successful formation of a bi-crystalline entity. The two nanocrystals may instead unbind and diffuse apart, or merge into a single larger nanocrystal, effectively contributing to a new process, $\Gamma_{1,1\to 1}$, not in the original model. Data fits [5,7,12] yield values of order $10^{-3}$ or smaller for $f$, which seems a rather drastic reduction.

Already at the level of the original "minimal" models, and one with the modification just described, numerical simulations sometimes require substantial computational resources. As a result, approximation techniques valid for the kinetics of larger clusters have been proposed [6] and then refined [12,13] and applied in the actual simulations, as well as compared to direct, full-model simulation results.

The second approach to modifying the model [7,8,66], uses a similar line of argument but in a somewhat different context. We point out that the "minimal" model already assumes a certain "bottleneck" for particle merger, by allowing only singlet capture. Indeed, the rates in Equation (23) with both $s_1 > 1$ and $s_2 > 1$, are all set to zero. This assumption was made based on empirical experimental observations that larger particles were never seen to pair-wise "merge" in solution. The conjecture has been that the restructuring processes that cause the observed rapid compactification of the growing secondary particles, and which are presently not understood experimentally or theoretically, mediate the incorporation of primary particles, but not larger aggregates, in the evolving structure, while retaining their crystalline core, to yield the final polycrystalline colloids.

It has been suggested that, small aggregates, up to certain sizes, $s_{max} > 1$, can also be dynamically rapidly incorporated into larger aggregates on diffusional encounters. Thus, we can generalize the model equations, see [7,8] for details, to allow for cluster-cluster aggregation with rates given by Equation (23), but only as long as at least one of the sizes, $s_1$ or $s_2$ does not exceed a certain value $s_{max}$. The sharp cutoff is an approximation, but it offers the convenience of a single new adjustable parameter. Indeed, data fits for CdS and Au spherical particles, yield good quantitative agreement, exemplified in Figure 5, with values of $s_{max}$ ranging [7,8,66] from ~ 15 for Au, to ~ 25 for CdS. Interestingly, these values are not only intuitively reasonable as defining "small" aggregates, but they also fit well with the concept of the cutoff value $n_{th}$, discussed in Section 2, beyond which atomistic aggregates develop a well formed "bulk-like" core. Indeed, the only available numerical estimate of such a quantity in solution [34], for AgBr nano-aggregates, suggests that $n_{th}$ is comparable to or somewhat larger than ~ 18.

We also comment that added cluster-cluster aggregation at small sizes, offers earlier formation of the initial peak in the secondary-particle distribution, which later further grows by the fast-capture-of-singlets mechanism. However, the singlet-capture-only, the added bottleneck-factor, and also small-cluster-aggregation approaches are all just different versions of modifications of the rates as compared to the standard diffusional-transport-driven irreversible-capture expression for the aggregation rate constants, Equation (23).

Thus, ultimately, as more microscopic experimental information on the colloid and nanoparticles growth processes becomes available, general systems of coupled aggregation equations, with more than a single-parameter pattern of rate modification and perhaps with the identification of pathways dependent not only on the aggregate sizes but also on their shape, morphology, and surface properties, should be developed. Finally, we note that allowing for cluster-cluster aggregation has required large-scale numerical effort and consideration of efficient algorithmic techniques for simulations, not reviewed here, including conversion of the

discrete-*s* equations to continuum ones, with the adaptive-grid (re)discretization both in the time, *t*, and cluster size, *s*, variables [7,8].

**5. Shape selection in particle synthesis**

Synthetic colloid and nanosize particles can assume many shapes and morphologies. Some particles are grown as single crystals. In other situations the growth does not yield a well-defined shape and structure. However, there is a large body of experimental evidence, e.g., [19,20,37,46,48,50,58,59,67-73] for growth of well-defined fixed-shape nonspherical particles under properly chosen conditions, even when they are internally polycrystalline. While semi-quantitative modeling of particle *size* selection has been successful, the challenge of explaining uniformity of *shape* and, more generally, *morphology* in many growth experiments, has remained largely unanswered until recently [14].

An exception have been the "imperfect-oriented attachment" mechanism [74-77] identified as persistency in successive nanocrystal attachment events leading to formation of uniform short chains of aggregated nanoparticles. Persistence can also mediate growth of other shapes [15,77] for a certain range of particle sizes. Indeed, nanosize and colloid particles for many growth conditions are simply not sufficiently large (do not contain enough constituent singlets) to develop unstable surfaces and/or the "dendritic instability" of growing side branches, then branches-on-branches, etc. — processes which distort a more or less uniform shape with approximately crystalline faces to cause it to evolve into a random or snowflake like morphology.

In modeling the *morphology* and *shape* selection, we have to consider several processes and their competition, which control the resulting particle structural features. In addition to diffusional transport followed by attachments of the atoms (ions, molecules) to form nanocrystals, or that of nanocrystalline building blocks to the growing particle surface to form colloids, these atoms/blocks can detach and reattach. They can also move and roll on the surface, as well as, for nanoparticles as building blocks, restructure and further grow diffusively by capturing solute species. Modeling all these processes presents a formidable numerical challenge.

Empirical experimental evidence [1] obtained primarily for spherical-particle colloids, have suggested that the arriving nanocrystals eventually get "cemented" in a dense structure, but retain their unique crystalline cores. Diffusional transport without such restructuring would yield a fractal structure [27,64]. More generally, however, on the experimental side, quantitative data on the time dependent kinetics have been rather limited. This represents a problem for modeling, because numerical results can only be compared to the measured distributions of the final particles and to limited data from their final-configuration structural analysis.

It is important to emphasize that particle synthesis processes are primarily carried out at large initial supersaturations which result in fast kinetics. Shape selection is then *not* that of the equilibrium crystal growth, even though the actual shapes frequently display properties of the crystallographic faces of the underlying material. One of the main difficulties in modeling particle shapes numerically [27,78,79], has been to describe the establishment of the crystalline (for nanoparticles) or compact (for colloids) stable "core" on top of which the growth of the structure then continues. Indeed, such a core is formed in the early stages of the growth, when a multicluster description is needed. At the later stages of the growth, the formed larger clusters are sufficiently dilute to treat each as a separate entity which grows by capturing matter (primarily singlets) from its environment.

Our kinetic Monte Carlo (MC) approach reported in [14], thus considers a seed, which is a pre-formed particle (nanocrystal or a few-singlet colloid precursor particle) which was assumed to be approximately spherical. This initial core captures diffusing "atoms" which can only be attached in positions locally defined by the lattice symmetry of the structure. This approach is motivated by nanocrystal growth, but can also shed some light on the formation of colloids, the faces of which follow the underlying symmetry presumably of a main singlet nanocrystal that dictates the orientation of the surface faces. Interestingly, for certain colloids, such as cubic-shaped polycrystalline neighborite ($NaMgF_3$), there are experimental observations [80] from dark-field and bright-field TEM, which can be interpreted as indicative of growth with crystal faces forming by the process of the outer shell of the particle recrystallizing itself to become effectively continuous single-crystal, on top of a polycrystalline core. Finally, in protein crystallization [81,82], the growth stage, from ~ $10^2$ to ~ $10^8$ molecules per crystal, after the initial small-cluster formation but before the onset of the really macroscopic growth modes, can also be analyzed by the single-core model.

This approach [14], while still requiring substantial numerical resources, has the flexibility of allowing to explicitly control the processes of particles (or atoms) "rolling" on the surface and detachment/reattachment, by using thermal-type, (free-)energy-barrier rules. The diffusional transport occurs in the three-dimensional (3D) space, without any lattice. However, the "registered" attachment rule starting from the seed, prevents the growing, moderate-size clusters from developing macroscopic defects and ensures the maintenance of the crystal symmetry imposed by the core. We can then focus on the emergence of the surface and shape morphological features. The results [14] have allowed to identify three regimes of particle growth.

The first regime corresponds to slow growth rates, for instance, when the concentration of externally supplied, diffusionally transported building blocks, to be termed "atoms," is low. In this case, the time scales of motion (hopping to neighbor sites and detachment/reattachment, which all can be effectively viewed as added no-surface diffusion) of the already attached atoms on the cluster surface, $\tau_d$, is much smaller than the time scale of the formation of new monolayers, $\tau_{layer}$. The shape of the growing cluster is then close to (but not identical with) the Wulff-construction configuration [83-86].

The second regime corresponds to fast growth, $\tau_{layer} \ll \tau_d$, and to the formation of instabilities of the growing cluster surface. The dynamics of the cluster shape is correlated with the spatial density of the local diffusional flux of atom intake. The flux is maximal near the highest-curvature regions of the surface. As a result, small-scale perturbations of the surface due to random fluctuations, are accompanied by increased diffusional flux of atoms to surface protrusions, which then further grow, provided that near such protrusions the influx of atoms overwhelms the outflow due to on-surface diffusion. Eventually the cluster assumes a form of a clump of sub-structures of smaller sizes.

The third regime, which is the most interesting for our study, corresponds to $\tau_d \sim \tau_{layer}$. This is a *nonequilibrium* growth mode, but, as demonstrated in [14], it can result in the particles developing and maintaining an even-shaped form, with well-defined faces that correspond to the underlying crystal structure imposed by the seed and by the attachment rules. This numerically found shape-selection was only obtained for a certain range of particle sizes. Thus, there is indeed the "persistence" effect alluded to earlier: As the particle grows larger, with more matter in it, growth modes involving bulges, dendritic structures, and other irregularities can be supported and are indeed realized.

The pattern of shape-selection in the nonequilibrium growth regime have been explored [14] for the simple cubic (SC), body-centered cubic (BCC), face-centered cubic (FCC), and hexagonal close-packed (HCP) crystal lattices. For instance, for the SC case, a cubic particle shape can only be grown in the nonequilibrium growth regime. There are several possible cluster shapes for a given type of crystal symmetry, the realization of each determined by the growth-process parameters.

Only a couple of illustrative results of an extensive numerical study [14] are presented below, for the 3D SC lattice. We first offer comments on the steady-state regime, followed by results for the nonequilibrium growth regime. Based on preliminary studies, the seed was defined by lattice cells within a sphere with radius of 15 lattice constants. The seed atoms were fully immobile. The latter assumption was made to save run time, based on observations that the seed in such simulations rarely evolved much from its original shape and density. Thus, only the atoms later adsorbed at the seed and the growing cluster, underwent the dynamical motion.

Let us first outline results for the "steady state" regime ($\tau_{layer} \gg \tau_d$) for the SC lattice. In this case, each atom attached to the cluster can have up to 6 bonds pointing to nearest neighbors, described by the set $\{\vec{e}_{int}\}$ of 6 lattice displacements of the type $(1,0,0)$. The set of displacements/detachments for surface atoms, $\{\vec{e}_{mov}\}$, was defined in two different ways. Case $A$: In this variant, $\{\vec{e}_{mov}\}_A$ included both the set $\{\vec{e}_{int}\}$ and also the 12 next-nearest-neighbor displacements of the type $(1,1,0)$, of length $\sqrt{2}$. Case $B$: here $\{\vec{e}_{mov}\}_B = \{\vec{e}_{int}\}$. In the latter variant, the dynamics of the surface atoms is slower.

Figure 6(a) illustrates the resulting steady-state particle shape for the variant $A$ of the SC simulation. We also show a schematic which illustrates the cluster shape formed with the type $(100)$, $(110)$, $(111)$ lattice planes, which happen to also be the dense-packed, low-index faces that dominate the low-temperature Wulff construction for the SC lattice [83-85]. However, this superficial similarity with the equilibrium Wulff shape is misleading. Indeed, our system dynamics does not correspond to thermal equilibration. The resulting shape is thus dependent on the dynamical rules. Specifically, Figure 6(b) shows the shape obtained for the same system but with variant *B* for the displacements/detachments, which imposes a slower surface dynamics.

This and other results lead to several interesting observations. First, the particle shape is not universal, even in steady state, in the sense expected [87] of many processes that yield macroscopic behavior in Statistical Mechanics: The microscopic details of the dynamical rules do matter. In practical terms this makes it unlikely that nonequilibrium particle shapes can be predicted based on arguments such as minimization of some free-energy like quantity. The second conclusion is that the surrounding medium can mediate processes that profoundly affect particle shape. The growth process should thus be considered in a self-consistent formulation that includes the particle's interactions with and the resulting transport of matter to and from its environment.

Another interesting observation is that well-defined particle shapes can be obtained in the present steady-state regime. Then why can't this regime be a candidate for predictable (within the present model) and well-defined particle shape selection mechanisms? The answer is in the observed [14] sensitivity the results to the density of and transport to and from the very dilute surrounding medium. Indeed, in this regime the isolated cluster assumption breaks down: Other clusters (particles) will compete for the "atoms" (solutes) in the dilute solution, and as a result growth mechanisms [26] that involve exchange of matter between clusters (Ostwald ripening) will become important.

Indeed, the main difference between the nonequilibrium and steady-state regimes is that the former corresponds to a fast, dominant growth process by capture of singlet matter from a dilute solution. Other processes, such as those involving exchange of matter with other clusters, or the on-surface diffusion, are slower. Thus, for nonequilibrium growth, the cluster shapes can be quite different. For example, for the SC lattice, a cubic shape, illustrated in Figure 7(a), was only found in the nonequilibrium regime [14] with the kinetic transition rates, detailed in [14], for atom intake vs. surface dynamics corresponding to $\tau_d \sim \tau_{layer}$. Other, less symmetrical shapes have also been found, as illustrated in Figure 7(b).

Examples of some regular shapes obtained for nonequilibrium growth with lattice symmetries other than SC, are given in Figure 8. A plethora of shapes obtained, for several lattices, is presented in [14], as are examples of unstable growth and other interesting growth modes, further exploration of which has been limited only by the demands of numerical resources required for simulating larger particles. We believe that the present model captures the key ingredients required for well-defined shape selection in the nonequilibrium growth regime. It avoids formation of macroscopically persisting defect structures. Then the dynamics of the growing particle's faces is not controlled by extended defects — which is a well known mechanism [82,86] that can determine growth modes in nonequilibrium crystallization. Apparently, this property allows the evolving surface to overwhelm imperfections, at least as long as the particles remain not too large, even for colloids that are formed from aggregating nanocrystalline subunits. The growing cluster faces then yield well-defined particle shapes and proportions. In fact, the densest-packed, low-index crystal-symmetry faces, which dominate the equilibrium crystal shapes, also emerge in the nonequilibrium regime. However, generally the particle shapes, planar faces and other surfaces present, and their proportions are not the same as in equilibrium.

## 6. Conclusion

We surveyed models and results for particle size and shape selection in colloid and nanoparticles synthesis. The reviewed theories, typically requiring numerical simulations to get results to compare with experiments or to gain qualitative insight into the model predictions, are presently limited and at best semi-quantitative. Furthermore, the experimental data are primarily limited to examination of the final products, whereas results for time dependent, kinetic processes, as well as detailed morphological data would be useful to advance our understanding of the fine-particle design, which would benefit diverse applications. Thus, notwithstanding the recent successes, we consider the status of the theoretical understanding of the kinetics of fine-particle synthesis, and the theory-experiment synergy as preliminary, and the field as facing interesting challenges and widely open for future research.

We wish to thank our colleagues D. Goia, I. Halaciuga, S. Libert, E. Matijević, D. Mozyrsky, J. Park, D. Robb and I. Sevonkaev for rewarding scientific interactions and collaboration, and acknowledge funding by the US ARO under grant W911NF-05-1-0339.

**FIGURES**

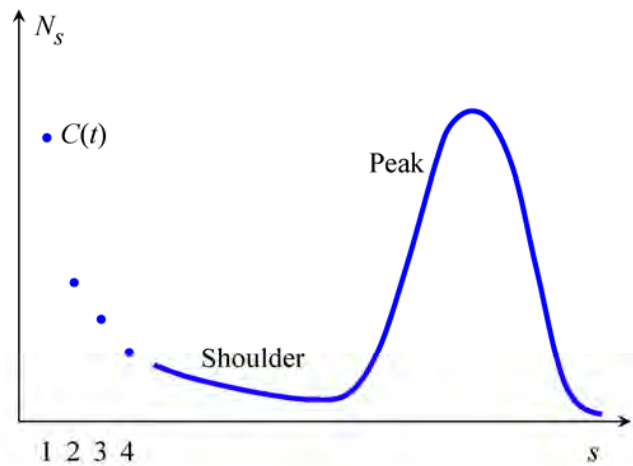

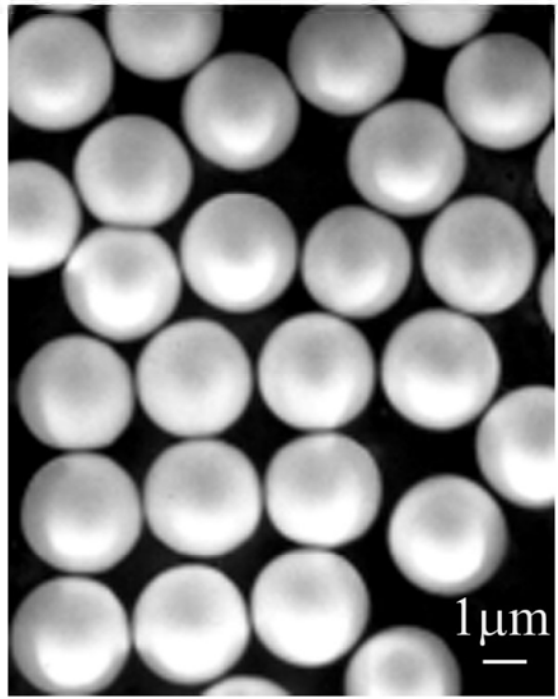

**Figure 1.** *Top:* the desired particle size distribution. The aim is to have the peak at the large cluster sizes develop by consuming singlets, by a kinetic process such that the width of the peak will remain relatively small. The points at $s = 1, 2, 3, 4$ emphasize that the $s$ values are actually discrete. *Bottom:* SEM image of polycrystalline spherical CdS colloid particles with uniform size and shape distribution.

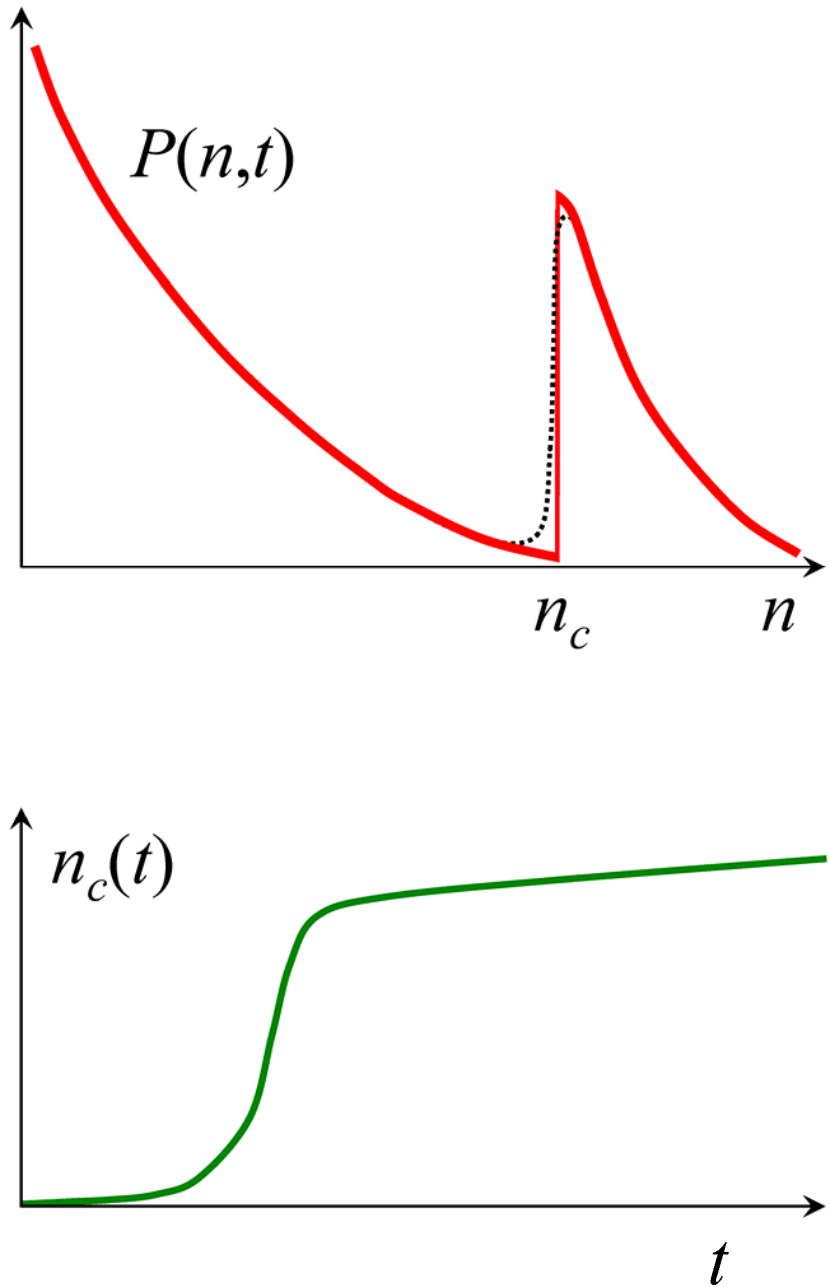

**Figure 2.** *Top:* Features of the large-time form of the cluster size distribution in the burst nucleation model approach. The actual distribution if steep but continuous near $n_c$, as indicated by the dotted line. *Bottom:* Time dependence of the critical cluster size. The induction period is followed by the "burst," and then the asymptotically linear growth (typically with small slope).

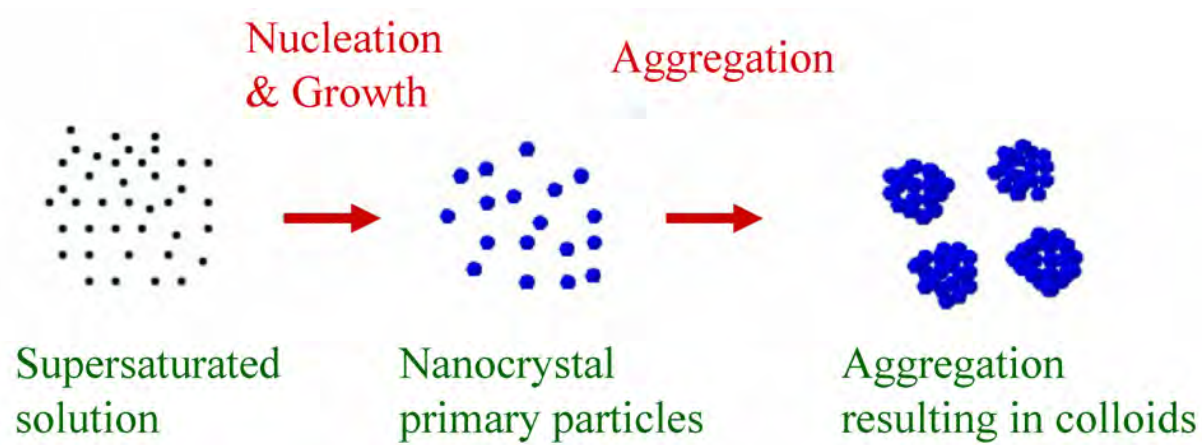

**Figure 3.** The two-stage mechanism for synthesis of uniform colloids by aggregation of nanocrystalline primary particles formed by burst nucleation in a supersaturated solution.

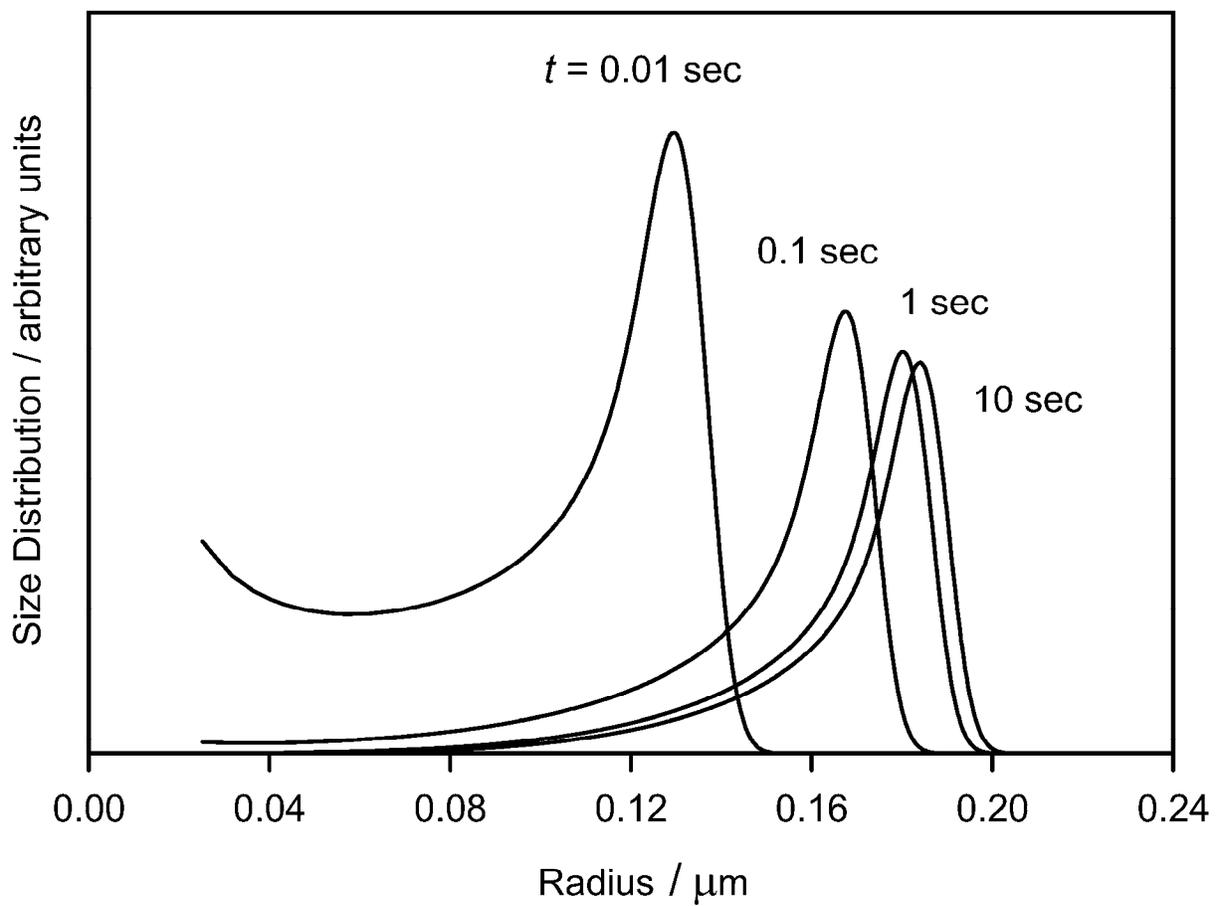

**Figure 4.** Example of a calculated colloid particle size distribution (in arbitrary units), plotted as a function of the colloid particle radius. The parameters used were for a model of formation of spherical Au colloid particles [1,5].

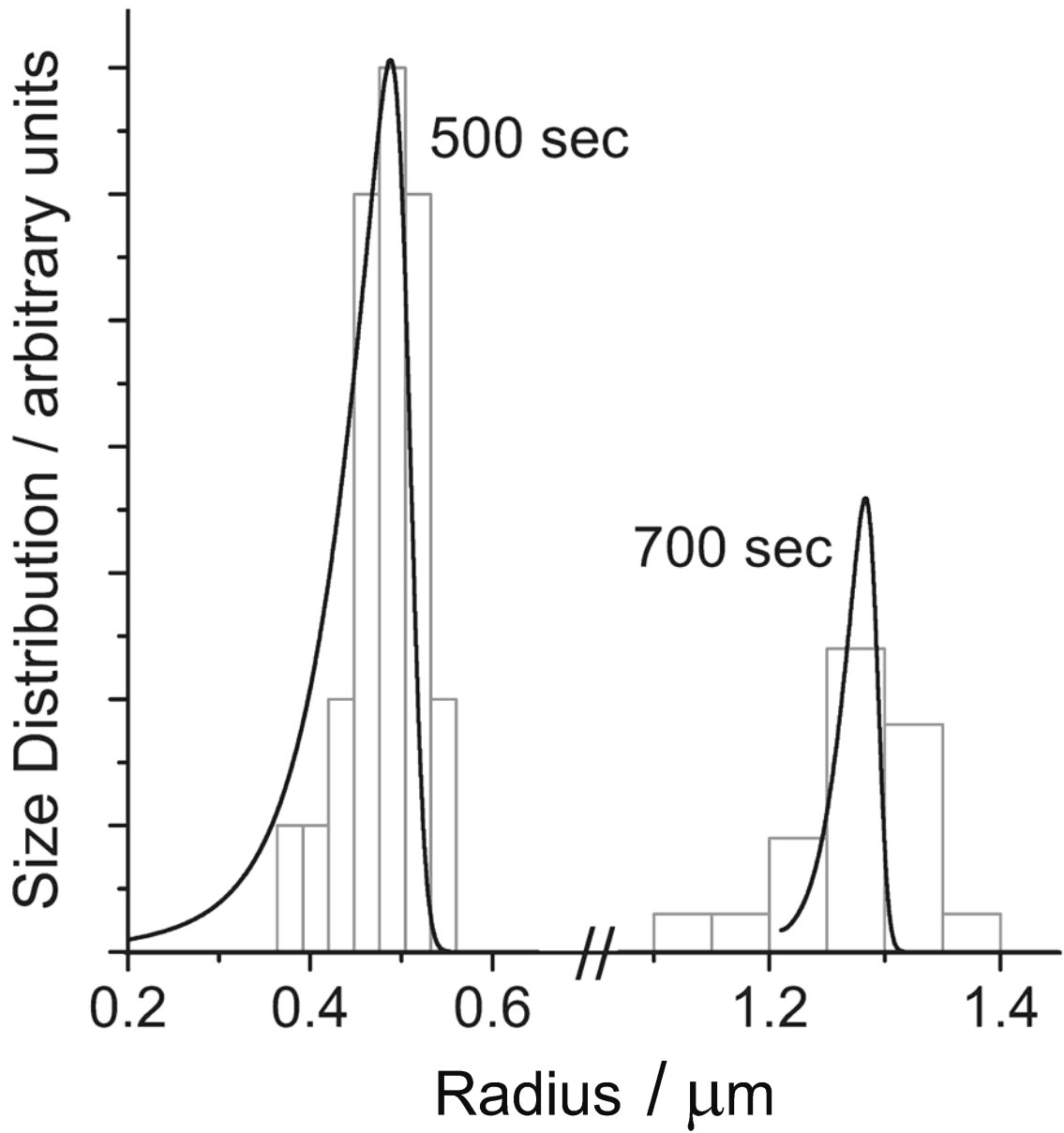

**Figure 5.** The calculated (curves) and experimentally measured (histograms) particle size distributions, for two different times during the growth. The parameters correspond to the $s_{max} = 25$ model [8] of formation of spherical CdS colloids.

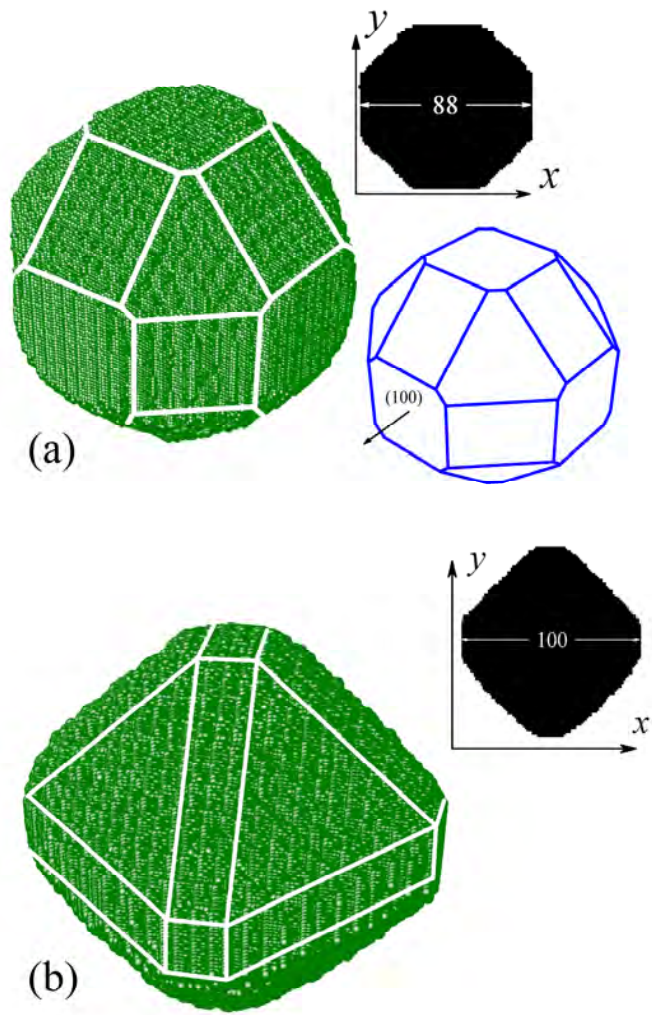

**Figure 6.** (a) Steady state SC lattice simulations for variant *A* of the displacements/detachments for surface atoms. The simulation details and parameter values are given in [14]. The resulting particle shape is shown for the cluster of $3.8 \times 10^5$ atoms, which was in a steady state with a dilute solution of free diffusing atoms. (The white lines were added at the edges to guide the eye.) Also shown is the projection of the cluster shape onto the *xy* plane, as well as the shape formed by lattice planes of the types (100), (110), (111) by an equilibrium Wulff construction (assuming that they all have equal interfacial free energies). (b) Steady state SC lattice simulations for variant *B* of the displacements/detachments for surface atoms. The dynamical rules were the same as for variant *A*. Also shown is the projection of the cluster shape onto the *xy* plane.

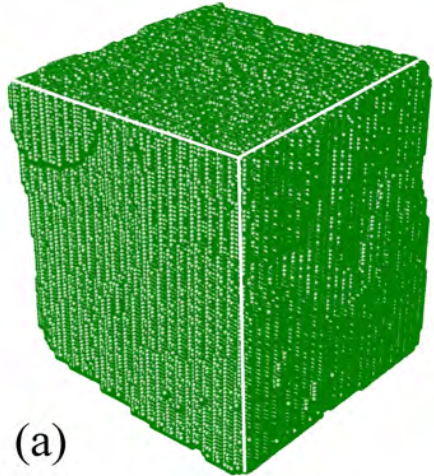

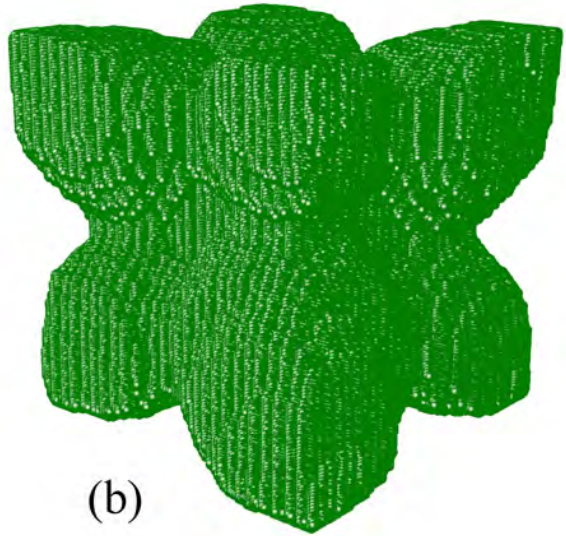

**Figure 7.** Examples of nonequilibrium SC lattice cluster shapes (for the kinetics of variant *A*). The parameters of the simulations, including the kinetics and the definition of the time units, are given in [14]. (a) The cubic shape emerges at rather short times, $t \approx 3 \times 10^5$, persisting for growing clusters, here shown for $t = 2.5 \times 10^6$, containing $4.5 \times 10^5$ atoms, with the cube edge length 77. (The white lines were added at the edges to guide the eye.) (b) Cluster grown with different parameter values, shown at time $t = 5.2 \times 10^6$, containing $1.8 \times 10^6$ atoms, of characteristic size 125.

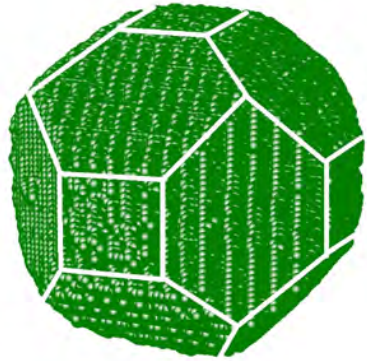

BCC

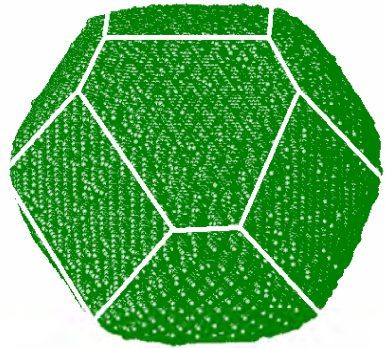

FCC

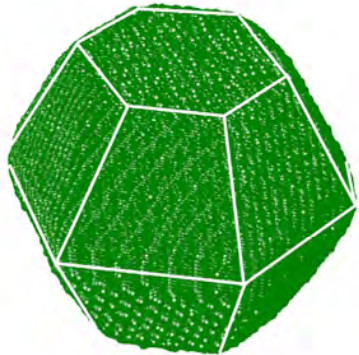

HCP

**Figure 8.** A selection of regular shapes obtained for nonequilibrium growth. The lattice symmetries are marked for each image. (The white lines were added at the edges to guide the eye.)